\documentclass[10pt, conference, letterpaper]{IEEEtran}

\usepackage{algpseudocode}
\usepackage{amsfonts}
\usepackage{url}

\usepackage{subfig}
\usepackage{graphicx}

\usepackage{epsf} 
\usepackage{pdfpages}
\usepackage{booktabs}
\usepackage{array}
\usepackage{subcaption}
\usepackage{verbatim}

\usepackage{amsmath}
\usepackage{amsthm}
\usepackage[numbers, sort]{natbib}

\newcommand{\etal}{\textit{et al.}\xspace}
\usepackage[ruled,linesnumbered]{algorithm2e}

\newcommand*{\nameone}{\textit{Scale}\xspace}

\usepackage{titlesec}
\usepackage{soul,color}

\def\BibTeX{{\rm B\kern-.05em{\sc i\kern-.025em b}\kern-.08em T\kern-.1667em\lower.7ex\hbox{E}\kern-.125em}}

\begin{document}

\title{\nameone: Deep Reinforcement Learning for Container Scheduling in Serverless Edge Computing}

\author{\IEEEauthorblockN{Chen Chen\IEEEauthorrefmark{1}, Zihan Jia\IEEEauthorrefmark{2}, Andrea Sabbioni\IEEEauthorrefmark{3}, Reza Farahani\IEEEauthorrefmark{4}, 
Lei Jiao\IEEEauthorrefmark{5}}\\
\IEEEauthorblockA{
        \IEEEauthorrefmark{1}Department of Computer Science, Nottingham Trent University, UK\\
        \IEEEauthorrefmark{2}Department of Computer Science, Loughborough University, UK\\
         \IEEEauthorrefmark{3}Department of Computer Science and Engineering, University of
Bologna, Italy\\
 \IEEEauthorrefmark{4}Distributed Systems Group (DSG),
TU Wien, Austria\\
 \IEEEauthorrefmark{5}Center for Cyber Security and Privacy,
University of Oregon, USA\\
		Email: chen.chen04@ntu.ac.uk; z.jia@lboro.ac.uk; 
andrea.sabbioni5@unibo.it; r.farahani@dsg.tuwien.ac.at; \\ljiao2@uoregon.edu}
}
    
\maketitle
\begin{abstract}
    Serverless computing has emerged as a promising computing paradigm for edge computing.
    However, adopting the event-driven model in highly dynamic, heterogeneous, and distributed edge systems poses significant challenges in request placement and resource management.
    Efficiently allocating requests to containers is therefore critical to reduce resource over-provisioning and unnecessary data movement.
    This paper proposes \nameone, a Service Level Objective (SLO)-aware container scheduling and resource allocation framework designed for serverless edge computing.
    \nameone employs a policy-based deep reinforcement learning algorithm to balance system stability and performance under dynamic workloads. The design jointly incorporates SLO constraints, end-to-end latency, and data locality into the scheduling decision process. Extensive simulations using large-scale real-world datasets from Huawei Cloud demonstrate that \nameone achieves solutions within a factor of 1.11–1.15 of a state-of-the-art Integer Linear Programming (ILP) solver, while reducing decision-making time by up to 99\%.
\end{abstract}

	\pagestyle{empty}  
	\thispagestyle{empty} 
\thispagestyle{empty}
\section{Introduction}

Serverless computing has emerged as a computing paradigm that abstracts infrastructure management and enables fine-grained, elastic resource provisioning, with Function-as-a-Service (FaaS) representing its most widely adopted execution model for event-driven workloads~\cite{10528903, Schmid2025,Wang2025}. By decoupling application logic from infrastructure concerns, serverless platforms relieve developers from explicitly provisioning, configuring, and managing compute resources, allowing them to focus primarily on application functionality. In this paradigm, developers submit function code to the FaaS platform, which dynamically provisions, scales, and manages the underlying execution environments in response to workload demand~\cite{10.1145/3578354.3592865, Huang2025}.
Serverless computing is particularly well-suited for applications with highly variable workloads, such as those in machine learning (ML) and the Internet of Things (IoT)~\cite{MohammadTariq2025}, where elastic scaling can significantly improve performance and resource efficiency. In addition, serverless platforms are commonly employed for processing data close to its source, including system operation logs and user-generated data, thereby reducing data movement and latency~\cite{Cui2024}.

Existing serverless platforms, such as Google Cloud Functions and AWS Lambda, allow developers to rapidly deploy a large number of container instances triggered by incoming events. To maintain a satisfactory end-user experience, many latency-sensitive applications enforce Service-Level Objectives (SLOs) on end-to-end latency. To meet these SLOs, service providers often rely on resource over-provisioning; however, this strategy may substantially increase overall resource consumption and adversely affect system performance.
Therefore, a key enabler of efficient resource allocation is SLO-aware container scheduling, which aims to continuously minimize user-perceived latency by assigning requests to the most latency-efficient containers while satisfying end-to-end SLO constraints. 

However, modern applications are increasingly moving from centralized architectures toward distributed edge deployments, enabled by advances in and the growing availability of edge computing infrastructures~\cite{11262634, farahani2025energyless}. As a result, user requests may be deployed across geographically distributed and heterogeneous edge nodes. This architectural shift introduces two key challenges: first, data movement must be considered due to the fact that the communication latency between the data source and host node is usually non-negligible. Second, the system must balance strict SLO guarantees against resource consumption in large-scale and dynamic deployments.

To address these challenges in serverless edge computing (\textit{SEC}), we decouple request scheduling from container reuse, enabling latency-aware execution while satisfying application-defined service-level objectives (SLOs). We formulate the container placement problem for serverless requests at the edge as an integer linear programming (ILP) model that captures fine-grained resource allocation decisions and request-level latency constraints, with the objective of minimizing end-to-end latency. Given the computational complexity of solving the ILP in large-scale or dynamic settings, we further design a deep reinforcement learning (DRL)–based approach. The proposed solution builds on actor–critic networks with a hierarchical action space, jointly considering system performance, workload dynamicity and SLO compliance.
The main contributions of this paper are listed as follows.
\begin{enumerate}
    \item We formulate an SLO-aware container scheduling problem in serverless edge computing, including comprehensive motivations, to minimize the end-to-end latency.
    \item We propose \nameone, a novel actor-critic network using a hierarchical action space to make decisions in two phases.
    The request scheduling is decided in the first phase, while the container reuse is decided in the second.
     The DRL agent is modeled to fully consider the temporal variability in workloads, the position of the request source and the position of the host node, ensuring a comprehensive understanding on the state and action space.
     
    \item We evaluate \nameone via large-scale simulations using real-world datasets and network topologies. By comparing with an ILP solver and another popular DRL approach, the results show that \nameone achieves latency within $1.15\times$ of the ILP, while reducing decision-making time by up to 99\%.
\end{enumerate}

\section{Related work}

Many recent works have extensively studied container scheduling in SEC, optimizing a number of objectives such as end-to-end latency, deployment cost, energy consumption and SLO violation rate.
A large amount of research has focused on mitigating the cold-start problem, as cold starts remain a major source of latency overhead in serverless systems. In particular, several studies~\cite{BasuRoy2024, Zhou2024, Yu2024a, Mahgoub2022, zhao2024taming} aim to reduce the frequency of cold starts through techniques such as container pre-warming, container reuse, and workload prediction.
Another direction is to reduce the cold-start time~\cite{Zhang2024, Kohli2024}. Different techniques are proposed to achieve this, such as partial code loading, container snapshot and checkpointing. In addition, other approaches focus on reducing the cold-start latency itself~\cite{Zhang2024, Kohli2024}. While the scale of serverless clusters is continuously increasing, a few works~\cite {Liu2025, Wang2025, farahani2024heftless, hou2023eavs} have focused on reducing the overhead introduced by serverless workflows and large-scale clusters, such as execution time.

However, those papers are largely orthogonal to \nameone, as they typically focus either on mitigating cold start or container placement in isolation, while overlooking the joint trade-off among end-to-end latency, SLO satisfaction, and container placement and reuse.

Yue~\etal~\cite{Yue2024} proposes a multi-agent reinforcement learning algorithm to co-optimize per-function placement and resource allocation.
The authors in~\cite{Yu2024} use DRL with an attention mechanism to harvest idle resources to accelerate serverless containers. Cui~\etal~\cite{Cui2024} uses DRL for serverless cluster upgrading.
Although existing DRL frameworks have shown promising performance in various settings, they are generally not designed to explicitly balance end-to-end latency objectives against constraints imposed by heterogeneous and limited resources. \nameone employs an actor–critic–based deep reinforcement learning architecture to achieve efficient container reuse and placement while maintaining acceptable SLO violation rates.

\section{Problem formulation}

\subsection{System model}
We model the edge infrastructure as a network $\mathcal{G} = (\mathcal{V}, \mathcal{E})$, which consists of $|\mathcal{V}|$ geographically distributed edge nodes. Each node $v \in \mathcal{V}$ is provisioned with finite CPU and memory capacities, denoted by $F_v$ and $C_v$, respectively. We use $B_v$ to denote the available bandwidth at edge node $v$. The link between the request source $s$ and its placement node $v$ is represented by $(s, v) \in \mathcal{E}$. Each request $k$ is characterized by a CPU requirement $f_k$ and a memory requirement $c_k$. For each request $k$, we denote by $s_k$ the container image size associated with its execution. Table~\ref{tab::var} summarizes all notations. 

\begin{table}[htb]
	\centering
	\small
	\caption{Symbols and Variables}
	\label{tab::var}
	\renewcommand\arraystretch{1}            
	\begin{tabular*}{250pt}{ll}
		\toprule	
		\textit{Symbols} & \textit{Description}\\
		\midrule
		$\mathcal{G} = (\mathcal{V}, \mathcal{E})$ & Physical network graph\\
            $\mathcal{V}$ & Set of edge nodes\\
            $\mathcal{E}$ & Set of links\\
            $\mathcal{K}$ & Set of requests\\
		$s_k$  & The image size of a container for request $k$  \\ 
        $B_v$  & The bandwidth of edge node $v$  \\ 
        $\alpha_k$  & The initialization time of a container 
        for request $k$  \\
              $c_k$ & The required amount of memory for request $k$\\
		$C_v$ & The memory capacity of edge node $v$\\
        $p_k$ & Required CPU time of request $k$\\
		$f_k$ & Required number of 
            vCPU cores of request $k$\\
        $F_v$ & The number of vCPU cores of edge node $v$\\
         $O_k$ & The SLO target for request $k$ \\
         \midrule
          \textit{Variables} &\\
          \midrule
		$x_{v}^{k}(t)$ & Binary variable equal to 1
            if request $k$ is assigned \\
            & to edge node $v$\\
        $z_v^k(t)$ & Binary variable equal to 1 if request $k$ needs 
        a \\& newly created container on node $v$\\
		\bottomrule
	\end{tabular*}
\end{table}

\subsection{Latency model}
We decompose the end-to-end latency into three distinct components, namely, the cold-start latency, computation latency, and communication latency.

\textit{(i) Cold-start latency} is incurred when initializing a new container by pulling images from remote repositories and preparing the execution environment. Thus, the cold-start latency is calculated as follows:
\begin{equation}
    T_{cold}^k =\sum\nolimits_{v \in \mathcal{V}} \bigg(\frac{s_k}{B_v} + \alpha_k \bigg) \cdot z_v^k
\end{equation}
where $s_k$ is the image size of a container requested by request $k$ and $B_v$ is the bandwidth of edge node $v$.
Thus, $\frac{s_k}{B_v}$ denote the time to pull the image from the remote repository.
$\alpha_k$ denotes the initialization time of a container requested by request $k$, which is assumed to be inversely proportional to the memory allocated to the request. In addition, $z_v^k$ is a binary variable indicating whether a container for request $k$ is instantiated on edge node $v$. Specifically, if a new container for request $k$ is created on node $v$, $z_v^k = 1$, and $z_v^k = 0$ otherwise.

\textit{(ii) Computation latency} is incurred when a container executes a request on an edge node. The computation latency depends on the computational demand of the request and the number of requested vCPU cores, and is modeled as follows:
\begin{equation}
    T_{comp}^k = \frac{p_k}{f_k} \cdot x_v^k,
\end{equation}
where $p_k$ is the CPU time required by request $k$, and $f_k$ is the number of vCPU cores required by request $k$. $x_v^k$ is a binary variable to decide whether request $k$ is hosted on node $v$.

\textit{(iii) Communication latency} is incurred when transmitting request $k$ from its source to the selected edge node $v$. It is mainly determined by the size of the request data and the available network bandwidth, and is defined as follows:
\begin{equation}
    T_{comm}^k = \frac{d_k}{\epsilon_{s,v}} \cdot x_v^k.
\end{equation}
where $d_k$ denotes the size of data required to execute a request, including input data and configurations, while $\epsilon_{s,v}$ denotes the link transmission rate from the source $s$ to node $v$.
Similar to prior work~\cite{Chen2018, Du2018}, we ignore the return communication latency, as it is typically negligible compared to the task execution time.
\subsection{Problem formulation}
We formulate the container scheduling problem for serverless computing as an optimization problem that minimizes end-to-end latency while guaranteeing application-defined SLOs under resource allocation constraints. Specifically, the objective is to minimize the aggregated end-to-end latency across the set of requests $\mathcal{K}$, subject to SLO satisfaction and resource capacity constraints.

\begin{equation}
\begin{aligned}
\min \quad & T = \sum\nolimits_{k \in \mathcal{K}} \left( T_{cold}^k + T_{comp}^k + T_{comm}^k \right) \\
\text{s.t.}\quad & \text{constraints in Eqs.}~(\ref{con::integer})-(\ref{con::slo}), \\
& \text{variables}~z_v^k, x_v^k \in [0,1].
\label{obj}
\end{aligned}
\end{equation}

First, we ensure that each request is assigned to exactly one edge node using Eq.~(\ref{con::integer}):
\begin{equation}
   \sum\nolimits_{v \in \mathcal{V}} x_v^k = 1, \quad \forall k \in \mathcal{K},
   \label{con::integer}
\end{equation}

We then guarantee that the total CPU demand of requests assigned to an edge node does not exceed its available CPU capacity using Eq.~(\ref{con::cpu}):
\begin{equation}
    \sum\nolimits_{k \in \mathcal{K}} f_k \cdot x_v^k \leq F_v, \quad \forall v \in \mathcal{V},
    \label{con::cpu}
\end{equation}

Similarly, Eq.~(\ref{con::mem}) ensures that the total memory demand of requests assigned to an edge node does not exceed the node’s memory capacity:
\begin{equation}
    \sum\nolimits_{k \in \mathcal{K}} c_k \cdot x_v^k \leq C_v,\quad  \forall v \in \mathcal{V},
    \label{con::mem}
\end{equation}



Finally, Eq.~(\ref{con::slo}) ensures that the end-to-end latency of each request does not exceed its SLO requirement $O_k$:
\begin{equation}
    T^k \leq O_k, \quad \forall k \in \mathcal{K}.
    \label{con::slo}
\end{equation}
\section{Deep Reinforcement learning Design}
\subsection{DRL model}
We develop an adaptive container scheduling policy for multi-tenant serverless edge computing. We present the key aspects of the DRL model as follows.

\textit{(i) State space:}
The system state is denoted via a one-dimensional vector, where the first segment describes the edge cluster specifications, i.e., [$U_v(t)$, $\omega_v^k(t)$, $\mathcal{K}$].
Here, $U_v(t)$ refers to the current CPU and memory capacity at edge node $v$ at time $t$, while $\omega_v^k(t)$ denotes the number and status (i.e., \textit{busy} or \textit{idle}) of existing containers at edge node $v$. $\mathcal{K}$ refers to the set of requests, where each request includes the source node of the request, the container it requires and the required hardware resources for the container.
The set of states $\mathcal{S}(t)$ can be formulated as follow:
\begin{equation}
    \mathcal{S}(t) = \{U_v(t),  \omega_v^k(t), \mathcal{K}\},
\end{equation}

\textit{(ii) Action space:}
We model the action as a multi-discrete action space, where $a_1(t)$ denotes the node to host the request, and $a_2(t)$ denotes whether request $k$ is assigned to a newly created container.
Note that $a_1(t)$ and $a_2(t)$ are correlated.
For example, if a node $v$ has no container available, then $a_2(t)$ must be true to create a new container for the new request.
We have embedded those correlations by creating internal constraints in the action space.
Thus, the action generated by our algorithm can be denoted as \mbox{$a(t)= [a_1(t), a_2(t)]$}.

\textit{(iii) Reward function:}
The reward function $r(t)$ is assigned to the agent at each step right after an action is selected.
To optimize our objective, the reward needs to resonate with the problem objective (Eq.~\ref{obj}).
The reward is a measurement to evaluate the performance of the action $a(t)$ under a given state $s(t)$.
For a given policy $\pi$, the reward assigned to the agent can be formulated as follows:
\begin{equation}
    r(t) = - \bigg( T_{cold}^k + T_{comp}^k + T_{comm}^k \bigg),
\end{equation}
where the negative sign encourages the agent to minimize the end-to-end latency. The goal of an agent is to optimize the cumulative rewards $R(t)$.
Therefore, the expected discounted reward function is presented as follows.
\begin{equation}
   R(t) = \mathbb{E}\left[\sum^{\infty}_{j=0} \gamma^j \cdot r(t+j)\right]. 
\end{equation}
where \mbox{$\gamma^j\in [0,1]$} denotes the discount factor to adjust the tradeoff between immediate and long-term returns. $j$ refers to the time steps.
For example, if it is close to 1 means the agent prefers long-term return rather than immediate return.
\subsubsection{Actor-Critic-based container scheduling framework}
We use two neural networks, namely the actor and the critic network.
The actor represents the policy network trained via gradient-based optimization, whereas the critic is a value-function network that assesses the quality of the policy produced by the actor.
We have used the proximal policy optimization (PPO) as the policy optimization implementation in the actor network.
The rationale is that PPO is proven to be stable as it learns a stochastic policy $\pi_\theta$ by using a clipping function.
In particular, \nameone updates its policy at the $k_{th}$ episode by:
\begin{equation}
    \theta_{k+1} = \arg \max_{\theta}  \mathbb{E} [\mathbb{L}(s(t), a(t), \theta_k, \theta)],
\end{equation}

where $\mathbb{L}$ refers the the surrogate advantage, which is a metric to measure how policy $\pi_\theta$ performs relative to the old policy $\pi_{\theta_k}$.
More specifically, we use the PPO-clipped version where $\mathbb{L}$ can be given as follows.

\begin{equation}
    \mathbb{L} = \mathbb{E}[\min (r_\theta(t) \cdot A(t), clip(r_\theta(t), 1-\epsilon, 1+\epsilon) \cdot A(t)],
\end{equation}

where $r_\theta(t) = \frac{\pi_{\theta}(a|s)}{\pi_{\theta_k}(a|s)}$ is the probability ratio between the old and the new policy. $\epsilon$ is a hyperparameter used to clip the objective function.
$A(t)$ is the advantage calculated as reward $r$ subtracted by baseline values.
The clip function is used to prevent excessive policy updates.
In other words, clipping keeps the new policy close to the old policy by constraining how much action probabilities can change.

The critic network predicts the state-value function $V_\phi(s(t))$ and learns by optimizing the difference between the target $(r(t)+\gamma V_\phi(s(t+1)))$ and the estimated values of the state, also known as the a one-step advantage esitmate as follows.

\begin{equation}
    J(\phi) = r(t) + \gamma V_\phi (s(t+1)) - V_\phi(s(t)),
\end{equation}

where $\gamma$ is the discount factor.
$ V_\phi (s(t+1))$ and $V_\phi(s(t))$ are the value predictions for next and current state.

To update the weights $\phi$ of the critic network: 
\begin{equation}
    \phi = \phi - \beta \cdot \nabla_\phi J(\phi).
\end{equation}

where $\nabla_\phi J(\phi)$ refers to the gradient of the network and $\beta$ refers to the learning rate for weights update.

\begin{algorithm}[htb]
  \caption{Actor-Critic based Container Scheduling Algorithm.}
  \label{algo::training}
  \small
  \SetKwInOut{Input}{Input}
  \SetKwInOut{Output}{Output}
Input: Initialize policy (actor network) parameters $\theta_0$ and value function (critic network) parameters $\phi_0$;~\label{code::init}\\
\For{episode $k$ $\leftarrow$ 0,1,2,...}
{   Collect set of trajectories $\mathbb{D}_k=\{\tau_i\}$ by running policy $\pi_k = \pi_{\theta_k}$ in the environment.
    Compute rewards-to-go $\hat{r}(t)$;\\
    Computer the estimated advantage $\mathcal{\hat{A}}$ based on the current value function $V_{\phi_k}$;\\
    Update the policy by optimizing the PPO-clip objective using stochastic gradient ascent:\\
    $\theta_{k+1} =  \arg \max_{\theta}\mathbb{E}[\mathbb{L}(s(t), a(t), \theta_k, \theta)]$;\\


    Fit value function by regression on mean squared error:\\

    $\phi_{k+1} = \arg \min_\phi \frac{1}{|\mathbb{D}_k|T}\sum_{\tau \in \mathbb{D}_k} \sum_{t \in T} (V_\phi(s(t)) - \hat{R(t)}) $
}
\end{algorithm}

\textbf{Algorithm}~\ref{algo::training} shows the pseudocode of the training process of the proposed actor-critic scheduling algorithm.
First, we initialize the actor and critic networks with random weights and set the hyperparameters (Line~\ref{code::init}).
At the beginning of each episode, the agent collects the trajectories $\mathbb{D}_k$, including actions, states and rewards.
Once an action is decided, the agent receives a reward and the state changes to the next.
At the end of each episode, the hyperparameters $\theta$ and $\phi$ are updated accordingly.

\section{Evaluation}
We implement \nameone with real-world traces in simulations using Stable-Baseline 3 with 2,000 lines of code in Python. The experiment is
conducted on a server with 32 GB RAM and a 13th Gen Intel(R) Core(TM) i7-13700H processor with 14 cores.

\subsection{Evaluation setup}
\textit{(i) Baselines.} To evaluate the simulation results of \nameone, we compare it with two benchmarks: \emph{(1) Midaco-solver~\cite{MIDACO}}, an Integer Linear Programming (ILP) solver that is widely used to approximate the optimum in optimization problems with high complexity. Midaco uses a derivative-free, evolutionary hybrid algorithm with an extension to mixed integer search domains. 
Constraints are handled within Midaco by the Oracle Penalty Method, which is an advanced and self-adaptive method to reach the global optimal solution.
Extensive numerical tests on hundreds of benchmarks~\cite{Schluter01072012} demonstrate its capability to obtain the global optimal solution fast and robustly on the majority of problems.
\emph{(2) m-DQN~\cite{Mampage2023}}, a multi-step value-based deep reinforcement learning algorithm that uses experience replay to stabilize training and uses a target network to reduce oscillations and divergence. We extend the algorithm to adapt to our optimization problem in serverless computing.

We implemented both m-DQN and \nameone in Stable-Baseline3, which provides a set of reliable implementations of reinforcement learning algorithms using PyTorch.

\begin{figure}[!htb]
    \centering
\subfloat{\includegraphics[width=0.35\textwidth]{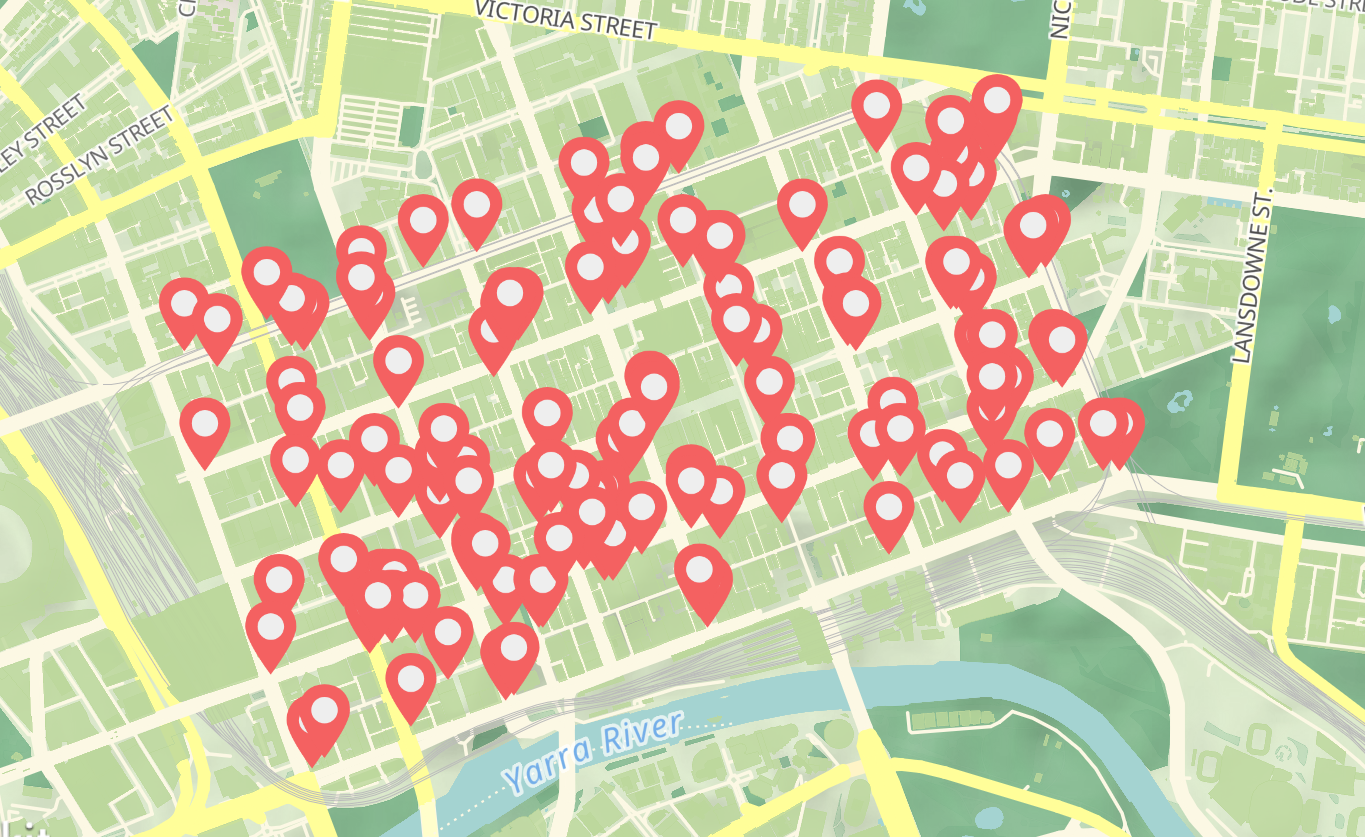}}
    \caption{Map of 125 edge nodes in Melbourne CBD area, the red dots represent the position of edge nodes.}
    \label{fig::topo}
\end{figure}

\textit{(ii) Edge network.}
We have used the edge network topology from the EUA dataset~\cite{Lai2018} that includes the location of 125 edge nodes in the Melbourne CBD area as shown in Figure~\ref{fig::topo}.
We defined 10 types of edge nodes with CPU frequency in [2.4, 3.6] GHz and memory capacity in [10, 30] GB.

\textit{(iii) Workloads.} We have used the Huawei dataset~\cite{huaweidataset} to generate the serverless requests, which includes the public invocations of Huawei clouds during 141 days with 200 functions. 
It is a cloud dataset due to the unavailability of serverless edge datasets.
To simulate the arrival patterns in edge computing, we build a request generator that uses the Zipf-$\beta$ distribution, creating workloads for each edge node.
This approach has been widely used to generate requests in edge computing~\cite{Shukla2018}.
We set the workloads' SLOs to be in [200, 400] ms, as reported in previous work~\cite{Bhasi2024}.

\subsection{Performance evaluation}
In the experiments, we report results of Midaco-solver with 50k, 100k, and 200k iterations.

\textit{(i) Overall performance analysis.}
\nameone yields a performance of end-to-end latency within a factor of 1.15 compared to the Midaco-solver, while the decision-making time is 99\% less.
The tradeoff is that \nameone efficiently approaches the results of Midaco-solver but saves crucial decision-making time.
Thus, \nameone is suitable for online, interactive, or latency-sensitive services, reducing task waiting time while efficiently approximating the results of Midaco-solver.

\begin{figure}[tbh]
    \centering
    \includegraphics[width=0.4\textwidth]{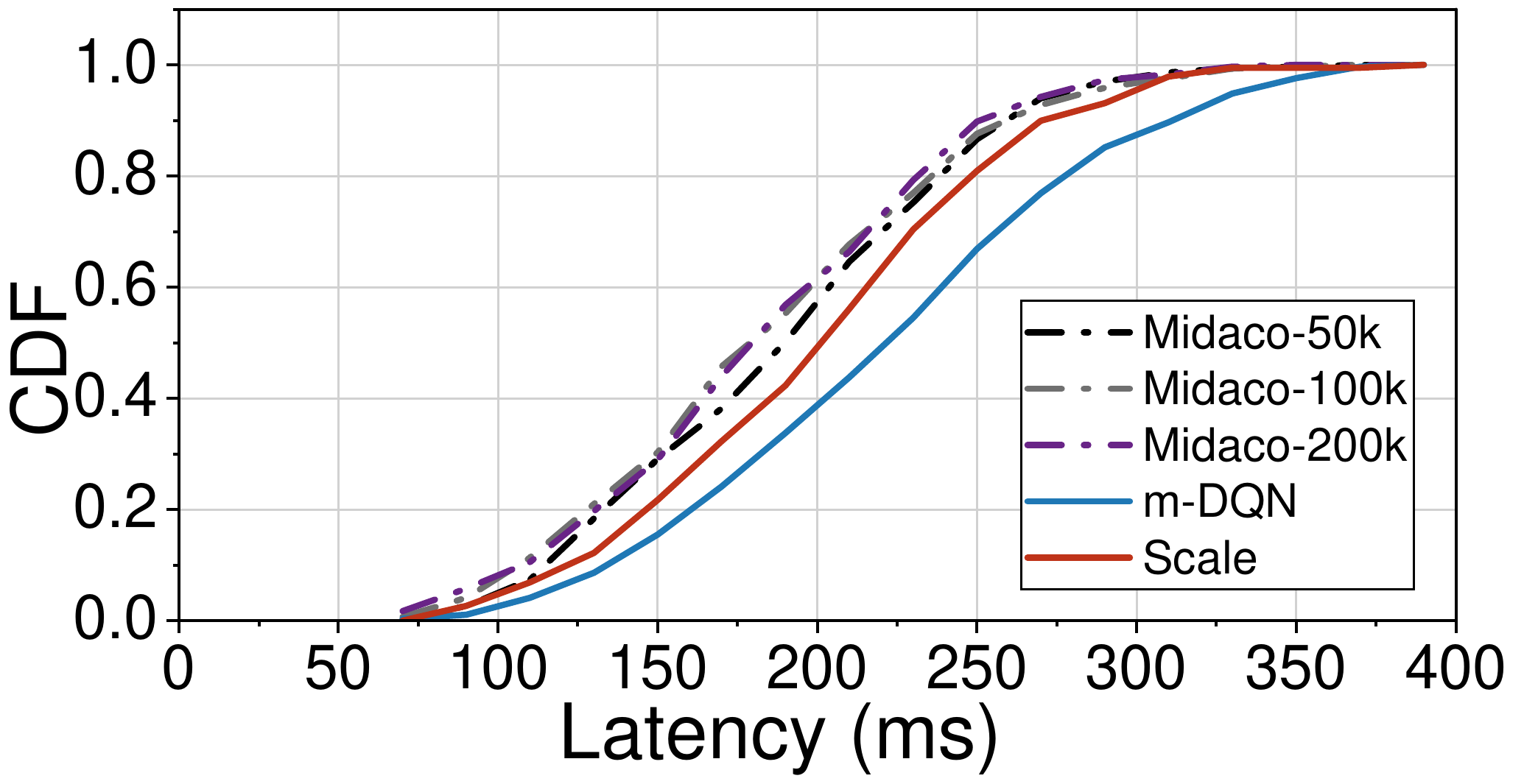}
    \caption{CDF of end-to-end latency.}
    \label{fig::e2e-latency}
\end{figure}

\textit{(ii) Distribution of end-to-end latency.}
Figure~\ref{fig::e2e-latency} shows the CDF distribution of end-to-end latency in all three approaches.
As discussed earlier, this end-to-end latency includes cold-start, computation, and communication latencies. As shown in Figure~\ref{fig::e2e-latency},
Midaco-200k receives the best performance at 284.41 ms for the 95th percentile of latency, while that of \nameone and m-DQN are 302.29 ms and 341.29 ms.
The average latency of Midaco decreases from 197.97 to 191.87 ms when the number of iterations increases from 50k to 200k; the performance is only improved by 3.08\%, indicating that Midaco is approximately converged at 200k iterations. 
The average end-to-end latency of \nameone is 208.93 ms. In contrast, the average latency of m-DQN is 320.82 ms.
This is because \nameone uses PPO, which adopts a clipped objective that prevents large, destabilizing policy updates, providing more stable performance for large-scale and dynamic environments.

\begin{figure}[htbp]
    \centering
    \includegraphics[width=0.4\textwidth]{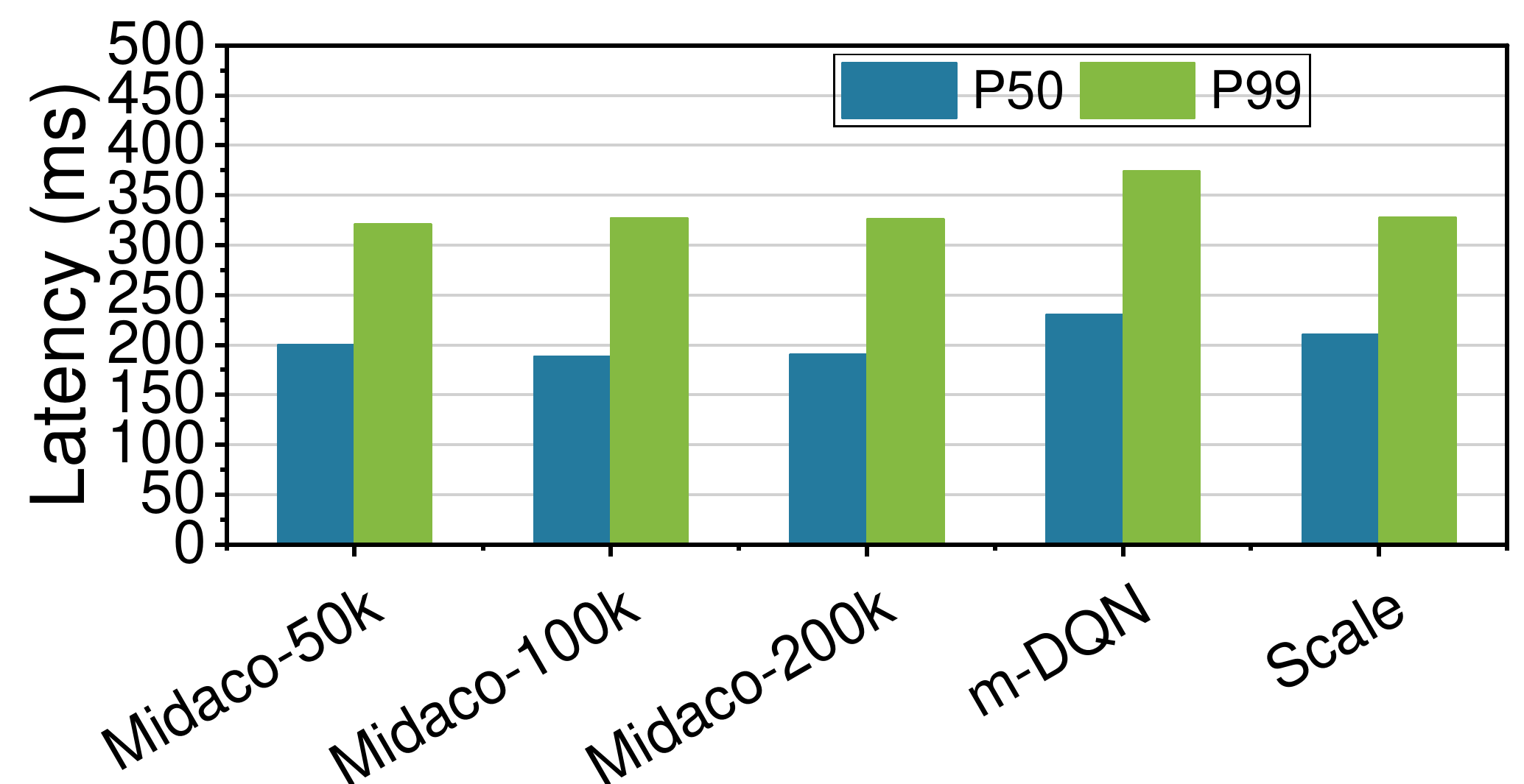}
    \caption{P50 and P99 latency.}
    \label{fig::p99}
     \vspace{-1mm}
\end{figure}

\textit{(iii) P50 and P99 latency.}
Figure~\ref{fig::p99} reports the P50 (50th percentile) and P99 (99th percentile) of latency for different approaches. P50 latency corresponds to the median request latency and reflects the typical user experience, whereas P99 latency represents the 99th-percentile latency and characterizes tail latency experienced by nearly all users.
We observe that the P50 latency of Midaco ranges from 188.42 to 199.71 ms while that of \nameone is 210.50 ms. 
In other words, the performance of \nameone is within a factor of 1.11 compared to the best case of Midaco, showing that \nameone efficiently approximates the results of Midaco.
Moreover, m-DQN obtains the worst performance at 230.63 ms because it suffers from instability due to overestimation bias and bootstrapping errors for large and dynamic environments.
We also notice similar trends in P99 latency, where \nameone efficiently approaches the results of Midaco within a factor of 1.15.

\begin{figure}[!t]
    \centering
    \includegraphics[width=0.4\textwidth]{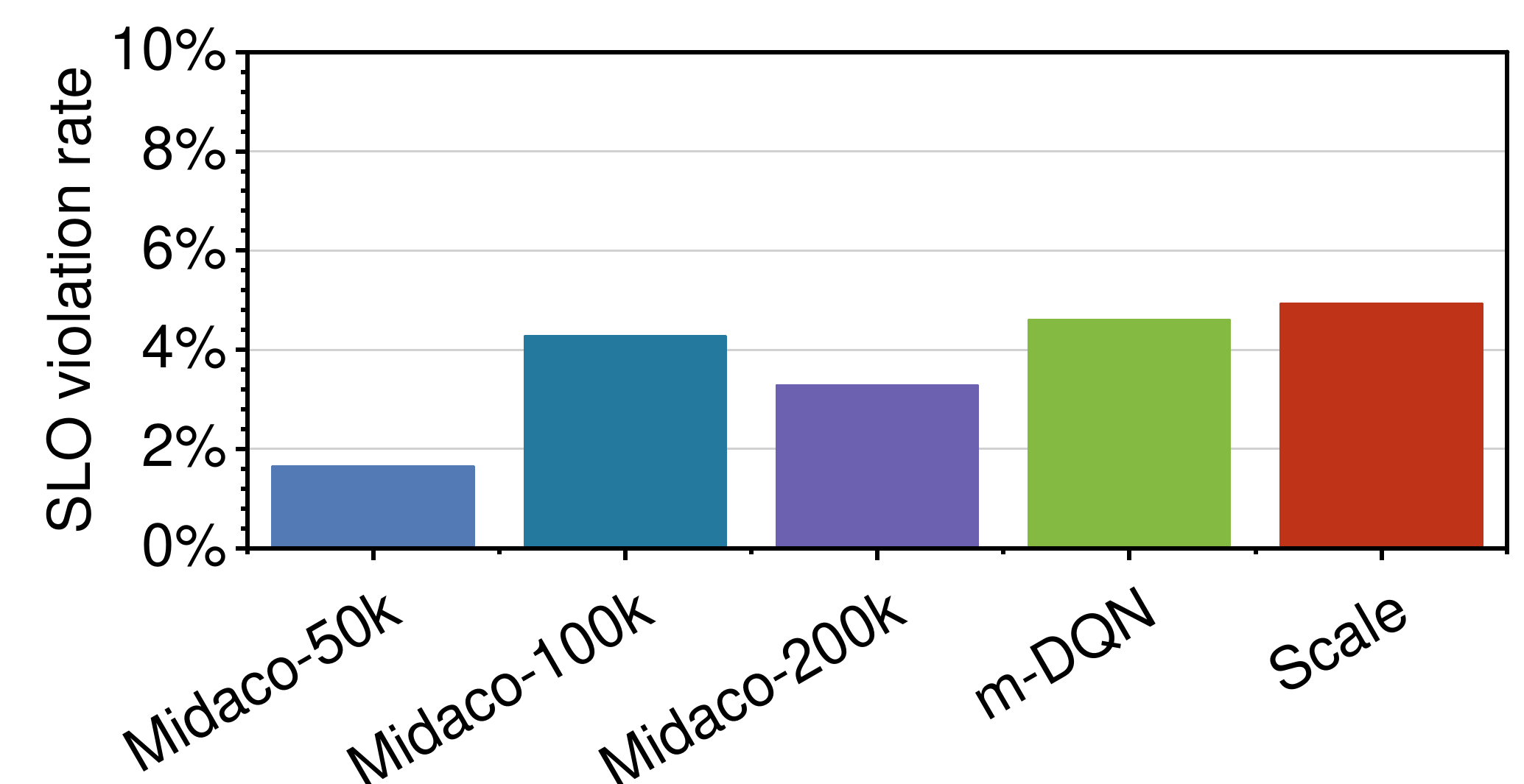}
    \caption{SLO violation rate in \%.}
    \label{fig::slo}
     \vspace{-1mm}
\end{figure}

\textit{(iv) SLO violation rate.}
The SLO violation rate is the ratio between the number of requests that violate the SLO and the total number of requests.
As illustrated in Figure~\ref{fig::slo}, \nameone yields 4.9\% while that of m-DQN is 4.6\%. The SLO violation rate of Midaco ranges from 1.6 to 4.2\%.

\begin{figure}[!t]
    \centering
    \includegraphics[width=0.45\textwidth]{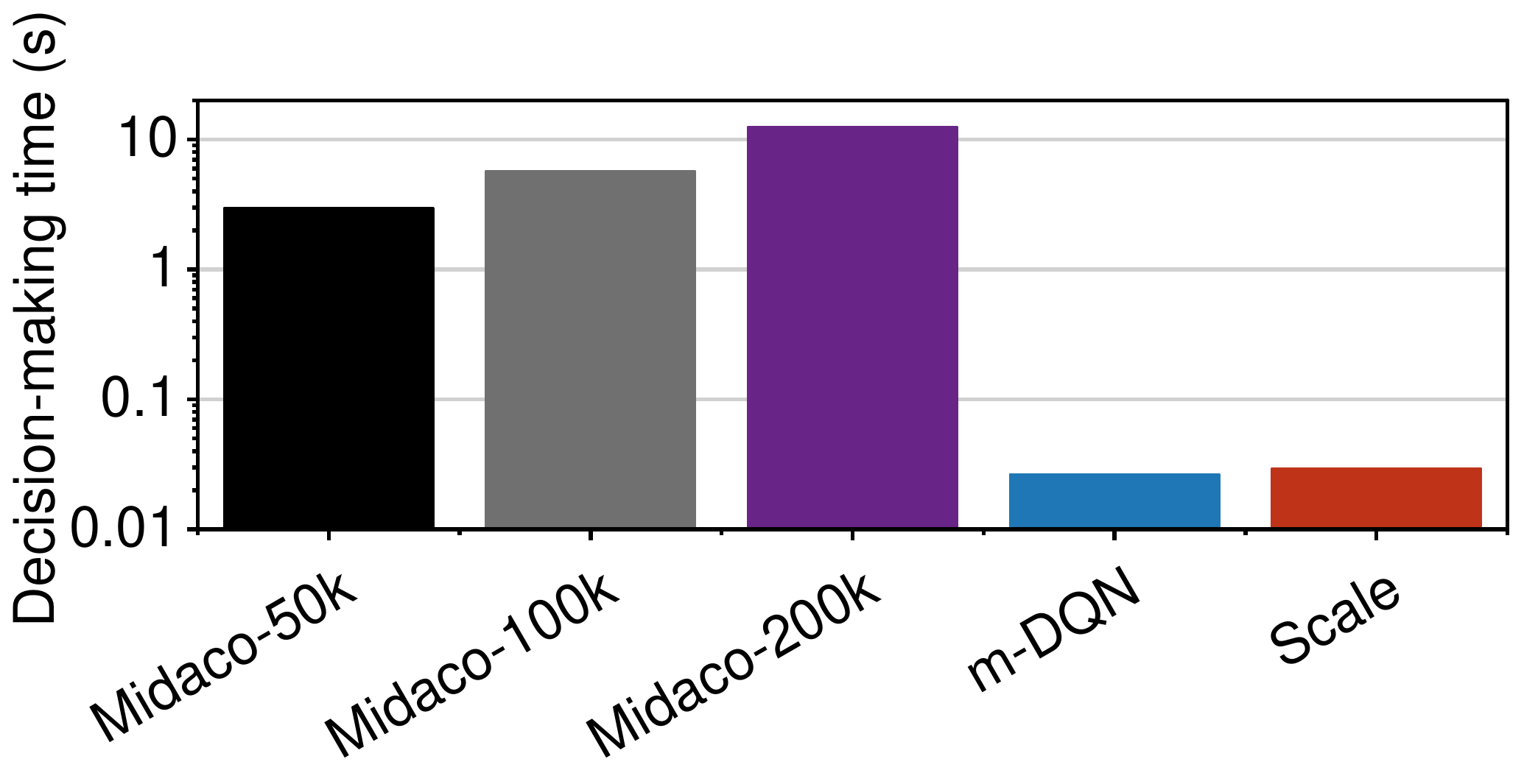}
    \caption{Decision-making time (s) per request.}
    \label{fig::d-time}
     \vspace{-1mm}
\end{figure}

\textit{(v) Decision-making time.}
Decision-making time reported in Figure~\ref{fig::d-time} refers to the time required by a scheduling approach to determine the placement of each incoming request. In edge environments, where resources are constrained and workloads are latency-sensitive, low decision-making time is critical to reducing request waiting time, enabling timely execution, and improving overall system responsiveness and user experience.
For \nameone and m-DQN, the decision-making time is the inference time per request. This is because we train them offline and use the inference in an online manner.
The decision-making time of \nameone and m-DQN are around 0.02 seconds per request, while that of Midaco ranges from 2.98 to 12.58 seconds with increased rounds of iterations.
The rationale is that Midaco needs to explore a large solution space before convergence, making it time-inefficient and unsuitable for online decision-making.
On the contrary, \nameone and m-DQN employ an agent to learn a near-optimal policy from the environment ahead of time in the training phase, enabling them to make decisions in an online manner.

\section{Conclusion}
This paper explored the container scheduling problem in serverless edge computing, aiming to minimize the end-to-end latency while meeting the SLO constraint for each request.
We proposed an actor-critic-based container scheduling framework to make efficient online decisions. We conducted extensive simulations using production datasets and real-world network topology.
\nameone obtains superior performance for the end-to-end latency within a factor of 1.11 to 1.15 compared to the LP solver Midaco.
Moreover, the decision-making time per request of \nameone is reduced by 99\% compared to Midaco Solver, showing that \nameone is suitable for online scheduling. As future work, we will extend \nameone to jointly optimize end-to-end latency and energy efficiency.

\section*{Acknowledgment}
This work was supported in part by the U.S. National Science Foundation under the grant CNS-2047719.

\bibliographystyle{unsrt}
{\footnotesize
\bibliography{reference-simplified}}


\end{document}